\documentclass[review]{elsarticle}

\usepackage{lineno,hyperref}
\usepackage{xcolor}
\usepackage{nicefrac}
\modulolinenumbers[5]

\usepackage{graphicx}
\usepackage{caption}
\usepackage{subcaption}
\usepackage{pdfpages}
\usepackage{amsmath}
\journal{Journal of \LaTeX\ Templates}
\begin{document}

\begin{frontmatter}

\title{A gaseous-helium cooling system for silicon detectors in the Nab experiment}

\author[utk]{Love Richburg}
\author[utk]{Noah Birge}
\author[utk]{Nadia Fomin}
\author[lanl]{Grant Riley}
\author[ornl]{Josh Pierce}
\author[ornl]{John Ramsey}
\author[ornl]{Wolfgang Schreyer}
\author[ornl]{Seppo Penttila}
\author[utk]{Isaiah Wallace}
\author[utk]{Di'Arra Mostella}
\author[uk]{Himal Acharya}
\author[asu]{Ricardo Alarcon}
\author[lanl,drex]{Ariella Atencio}
\author[ornl]{Leah Broussard}
\author[ncsu]{Jin Ha Choi}
\author[asu]{Skylar Clymer}
\author[uk]{Christopher Crawford}
\author[lanl]{Deion Fellers}
\author[eku]{Jason Fry}
\author[lanl,chi]{Duncan Fuehne}
\author[utk]{Zachary Garman}
\author[ornl]{Corey Gilbert}
\author[utk]{Rebecca Godri}
\author[ornl]{Francisco Gonzalez}
\author[utc]{Josh Hamblen}
\author[ornl]{Sean Hollander}
\author[gt]{Aaron Jezghani}
\author[uva]{Huangxing Li}
\author[uwin]{Nick Macsai}
\author[lanl]{Mark Makela}
\author[uwin]{Russell Mammei}
\author[ornl]{David Mathews}
\author[lanl]{Pat McGaughey}
\author[uman]{August Mendelsohn}
\author[lanl]{Jackie Mirabal}
\author[uk]{Austin Nelsen}
\author[utk]{Jordan O'Kronley}
\author[utk]{Hunter Presley}
\author[asu]{Glenn Randall}
\author[uva]{Americo Salas-Bacci}
\author[ornl]{Alexander Saunders}
\author[lanl]{Erick Smith}
\author[ncsu,lanl]{Bryan Zeck}

\address[utk]{University of Tennessee, Knoxville}
\address[ornl]{Oak Ridge National Laboratory}
\address[lanl]{Los Alamos National Laboratory}
\address[asu]{Arizona State University}
\address[eku]{Eastern Kentucky University}
\address[gt]{Georgia Tech}
\address[ncsu]{North Carolina State University}
\address[uk]{University of Kentucky}
\address[uman]{University of Manitoba}
\address[uwin]{University of Winnipeg}
\address[umi]{University of Michigan}
\address[uva]{University of Virginia}
\address[utc]{University of Tennessee, Chattanooga}
\address[mit]{Massachusetts Institute of Technology}
\address[lpc]{Laboratoire de Physique Corpusculaire de CAEN}
\address[drex]{Drexel University}
\address[chi]{University of Chicago}

\begin{abstract}
The Nab experiment aims to extract the neutron beta decay correlation coefficients `a' and `b'. This will be accomplished using a 7 m tall electromagnetic spectrometer which measures electron energies and proton momenta. Detection of electrons and protons resulting from neutron beta decay will be carried out using large-area, thick, highly-segmented, single-crystal silicon detectors. These detectors and accompanying electronics will be cooled by a recirculating, gaseous helium cooling system to below 150 K with $\pm$ 0.5 K stability. We will motivate the need for detector cooling in the Nab experiment and discuss design and performance of this cooling system.
\end{abstract}

\end{frontmatter}

\section{Introduction}

Free neutron beta decay is a useful tool to study aspects of the weak force described by the Standard Model of Particle Physics (SM) \cite{Dubbers_2011}. Measurements of neutron lifetime and beta decay correlation coefficients are used to constrain SM parameters and probe possible Beyond Standard Model (BSM) physics. One of the aspects of the weak force that is currently of particular interest is the unitarity of the Cabbibo-Kobayashi-Maskawa (CKM) quark mixing matrix \cite{Aoki_2020, Falkowski_2021}. As the largest element in the first row of the CKM matrix, $V_{ud}$ makes a significant contribution to the unitarity test. The inputs required to determine $V_{ud}$ from free neutron beta decay are the neutron lifetime and the ratio of the axial-vector to vector coupling constants, $\lambda = g_A/g_V$ \cite{PDGlive}. Two of the beta decay correlation coefficients, $A$ (the beta decay asymmetry parameter) and $a$ (the electron-antineutrino correlation coefficient), are suitably sensitive to variations in $\lambda$ to motivate using them to access it \cite{Dubbers_2011}. There have been many previous determinations of $\lambda$ from $A$ which are nearly in agreement with each other \cite{PDGlive}. Recently, there has been a push to access $\lambda$ using $a$ to provide a determination of $\lambda$ that is subject to a different set of systematics. Two recently updated determinations by aSPECT (2024) \cite{PhysRevLett.132.102501} and aCORN (2024) \cite{Wietfeldt:2023mdb} of $\lambda$ from $a$ are consistent with each other, but not consistent with $A$ determinations. 

\begin{table}[h!]
\caption[Comparison of most updated $a$ measurements from aSPECT and aCORN to Nab projection]{Comparison of most updated $a$ measurements from aSPECT and aCORN to Nab projection}
\begin{center}
\begin{tabular}{ |c|c|c|c|c| } 
 \hline
 Experiment Name & Year & $a$ & Uncertainty & $\Delta a/a$\\ 
 \hline
 aSPECT & 2024 & -0.10402 & $\pm 0.00082$ & $7.9 \times 10^{-3}$ \\ 
 \hline
 aCORN & 2024 & -0.10779 & $\pm 0.00183$ & $1.7 \times 10^{-2}$ \\ 
 \hline
 Nab (projected) & - & - & - & $\sim 1 \times 10^{-3}$\\
 \hline
\end{tabular}
\end{center}
\label{tab:asummary}
\end{table}

The first aim of the Nab experiment is to determine $a$ with relative uncertainty $\Delta a/a \sim 1 \times 10^{-3}$. This will be the most precise determination of $a$ that has ever been made from free neutron beta decay. The extracted value of $a$ will be used to extract $\lambda$ and subsequently $V_{ud}$ to test the unitarity of the CKM matrix \cite{Nab:2012hnc,Fry,Nab:2008cwh}. The most precise assessments of $V_{ud}$ have come from super-allowed Fermi transitions \cite{PDG22} which require nuclear corrections. With Nab's projected precision in $a$ and next-generation neutron lifetime measurements, extractions of $V_{ud}$ via the free neutron will be competitive with super-allowed Fermi transitions without needing nuclear corrections.

The secondary goal of Nab is to determine the Fierz interference term, $b$, to $\Delta b \approx 3 \times 10^{-3}$ for the free neutron from the decay electron energy spectrum. The Fierz interference term vanishes in the SM. A nonzero value of $b$ would be indicative of BSM physics \cite{Burgess_Moore_2006}.

\subsection{The Nab Experiment}

\begin{figure}[htb]
  \centering
  \includegraphics[width=0.50\textwidth]{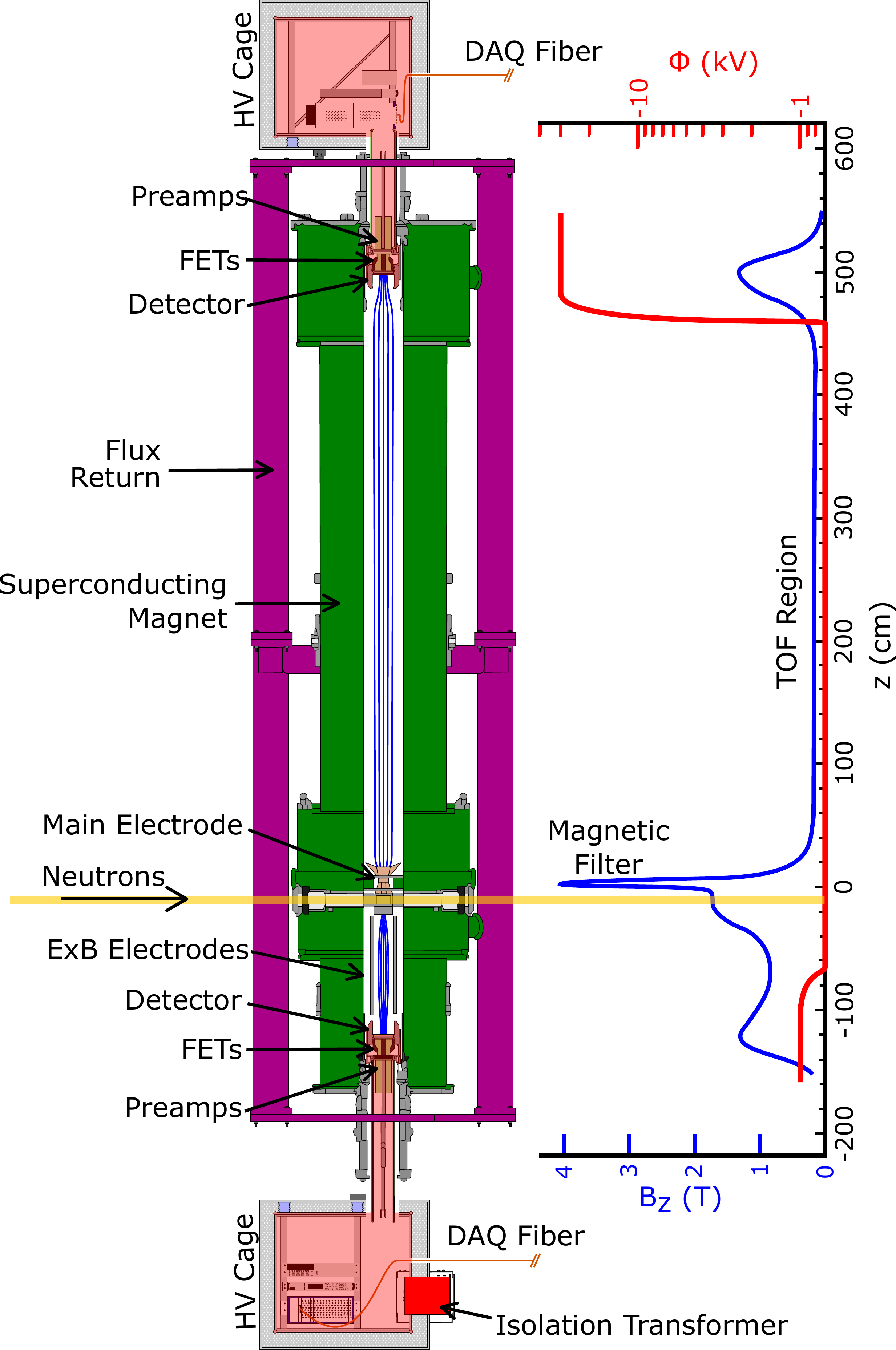}
  \caption{Shown above is a schematic of the Nab spectrometer, with the electric (red) and magnetic field (blue) profiles indicated in the plot to the right. \cite{Fry}.}\label{fig:nab_model}
\end{figure}

The Nab experiment apparatus is currently installed at the Fundamental Neutron Physics Beamline at the Spallation Neutron Source at Oak Ridge National Laboratory \cite{menu_proceedings}. Nab utilizes a 7 m tall cryogen-free superconducting electromagnetic spectrometer with two large-area, 2~mm thick, pixelated silicon detectors, one at each end. Detailed descriptions of the Nab spectrometer and detection system can be found in Refs. \cite{leendert,Broussard2017,Broussard2017a},
and a schematic of the Nab spectrometer can be seen in Fig.~\ref{fig:nab_model}. 

As the pulsed, unpolarized, cold neutron beam passes through the Nab spectrometer, a portion of the neutrons decay within the fiducial volume. The decay products (protons and electrons) are guided to the upper and lower detectors by the magnetic field. The Nab detectors are used to identify proton events and directly measure electron energies. Coincident protons and electrons will be identified to determine the proton time of flight (TOF) from the time difference. The Nab spectrometer is asymmetric to maximize the flight path of the protons to the upper detector and allow the carefully designed magnetic field to fully longitudinalize the proton momenta and minimize the TOF uncertainty. The upper detector is configured so that the protons pass through a -30~kV acceleration potential before reaching it (Fig. \ref{fig:nab_model}). This serves to increase the proton energies so that they can penetrate the dead layer of the detector. The Nab experiment extracts $a$ using proton momenta (calculated from the measured proton TOF) and the coincident electron energies.

\section{Detector cooling system design requirements and challenges}

Energy resolution, timing resolution, and detection efficiency are partially influenced by the temperature stability of two main components: (1) the detector and (2) the amplifying electronics:  
\begin{enumerate}
    \item With detector temperatures of 150~K or less, the signal-to-noise ratio is sufficient for seeing proton signals. Temperature stability is necessary to maintain a consistent detector response to incoming charged particles.
    \item Similarly, the BF862 N-channel junction field-effect transistors (FETs) used for initial signal amplification must be kept at a relatively constant temperature since their gain varies with temperature. They must also be cooled to mitigate the heat load from the amplifying electronics on the detector.
\end{enumerate}
Hence, the detectors and a portion of the accompanying front-end electronics must both be actively cooled. A recirculating, gaseous-helium-based cooling system has been designed to accomplish this. The low specific heat of helium offers fine control over system temperatures, and its low boiling point circumvents some of the complications that can arise with cryogenic liquid-based cooling. Details of the design and performance of this system will be discussed below.

The two main goals for the Nab detector cooling system are to cool both detectors to below 150~K and to achieve $\pm$~0.5~K stability. Cooling the detectors to below 150~K reduces noise due to thermally excited electron-hole pairs in the silicon since this process is positively correlated to temperature. As Knoll tells us, ``the probability per unit time that an electron-hole pair is thermally generated is given by'':

\begin{equation}
    p(T) = CT^{3/2}e^{-\frac{E_g}{2kT}},
\end{equation}
where $T$ is absolute temperature, $E_g$ is bandgap energy, $k$ is the Boltzmann constant, and $C$ is a proportionality constant characteristic of the material \cite{Knoll}. The reduction of this effect can be observed as a decrease in detector leakage current as the detector is cooled. Nab's silicon detectors are sensitive to temperature-dependent changes in electron hole pair creation energy \cite{leendert}. This temperature dependence means that variations in temperature can lead to different \textit{measured} energies for incident particles of the same energy. The mobility of charge in silicon is also temperature dependent. This directly affects rise times of signals \cite{leendert}. For Nab to meet its precision goals, the systematic offset in the average particle arrival times must be understood to about $\pm$ 0.3~ns. 

\begin{figure}[ht]
  \centering
  \includegraphics[width=1.0\textwidth]{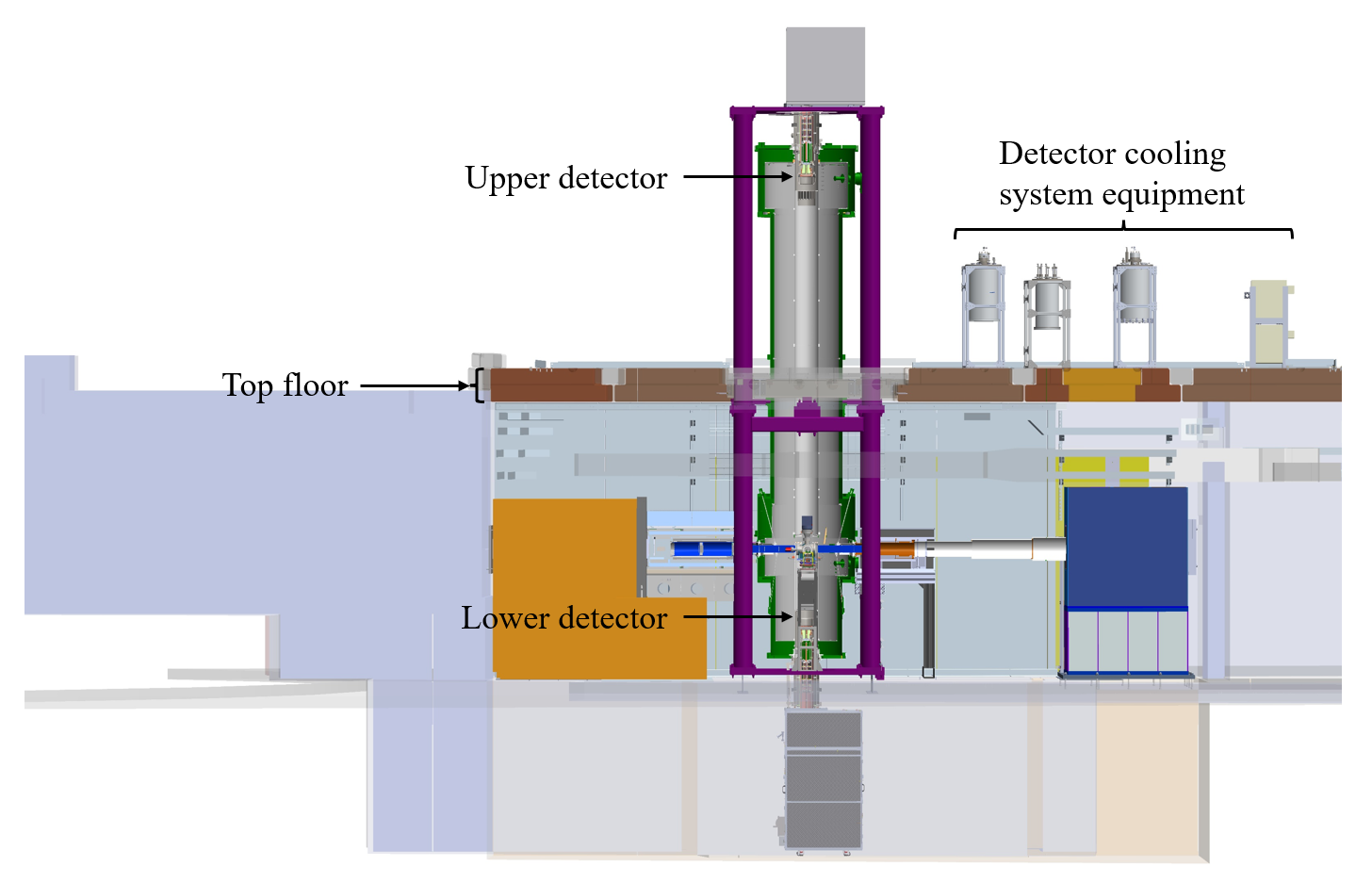}
    \caption{This schematic shows the locations of the detectors relative to the placement of the detector cooling system equipment.}
  \label{fig:view_1}
\end{figure}

The detectors are located at the upper and lower ends of the 7~m tall Nab spectrometer and there is limited floor space near the detectors. The cooling system had to be designed around these spatial constraints (see Fig. \ref{fig:view_1}). So, the bulk of the cooling equipment was installed on the more spacious ``top floor" of the beamline, roughly level with the middle of the magnet, and cold helium is delivered to each detector via long vacuum-insulated transfer lines. 

The detectors are operated at high voltage. In particular, for the measurement of $a$ the upper detector is floating at -30~kV (this has been discussed in Section 1.1). For safety, access to the detectors must be contained within Faraday cages. Thus, the insulated helium lines must traverse a high voltage break to reach the detector system. To do this, we designed specialty components that are electrically and thermally insulated as discussed in Section 3.

Data collection for Nab proceeds for months at a time. To best support this, the cooling system must be able to operate continuously for these long periods. Thus, the system is a closed loop that recirculates helium gas.

Since Nab is a high precision experiment, it is also crucial to minimize introducing any noise to the detectors which could obscure proton signals. In early tests with LN$_2$ as the coolant, disruptively high levels of noise were observed in the detectors. Noise due to physical shaking of the cooling lines and hence microphonics is minimized by using helium gas instead of liquid nitrogen. 

We are able to achieve cost-effective helium circulation by implementing two double diaphragm process pumps which must be operated above the freezing temperature of water instead of utilizing much more expensive pumps which are capable of circulating media at cryogenic temperatures. We recover some of the cooling power that is lost by warming the gas on each cycle with off the shelf counter flow heat exchangers \footnote{https://www.mcmaster.com/8546T11/}. This is discussed in Section 3.

\section{Nab detector cooling system design}

\begin{figure}[ht]
  \centering
  \includegraphics[width=1.0\textwidth]{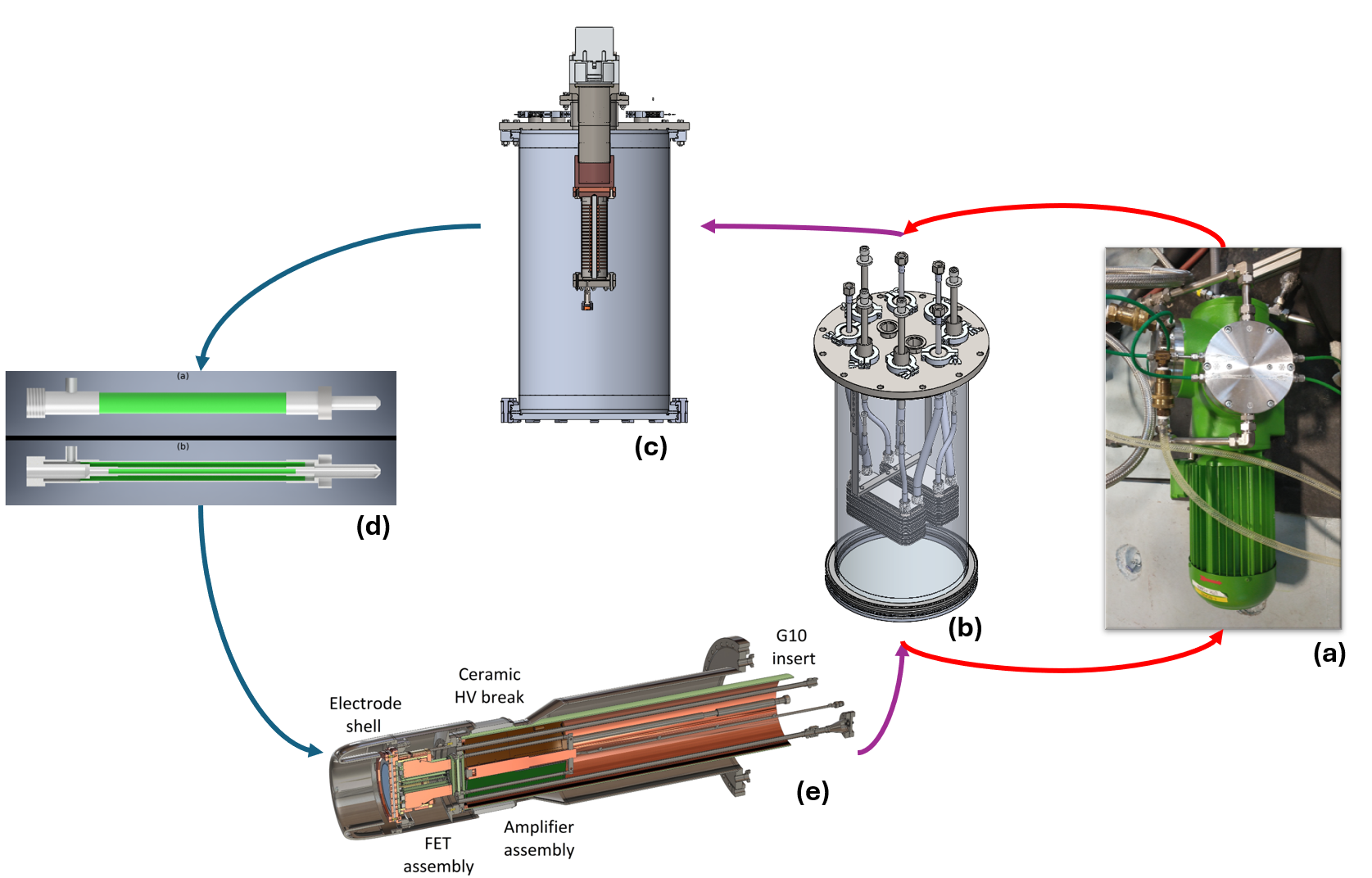}
    \caption{Representative flow diagram of one cooling loop (the second loop is functionally identical):(a) KNF Neuberger double diaphragm process pump, (b) Counter flow heat exchanger (``Tardigrade''); vacuum vessel and both counter flow heat exchangers are shown, only one is used in each loop, (c) Primary heat exchanger and vacuum vessel (``Cool-o-stat''), (d) Custom G-10 high voltage feed through (HVFT), (e) Nab detector system.}
  \label{fig:full_system}
\end{figure}

The Nab detector cooling system consists of two independent, helium gas recirculating, closed loops. One loop is for the upper detector system and one is for the lower detector system. The two loops are nearly identical with the only difference being in the geometry of the transfer line routes.

\begin{figure}[htb]
\centering
\begin{subfigure}{.5\textwidth}
  \centering
  \includegraphics[width=5cm]{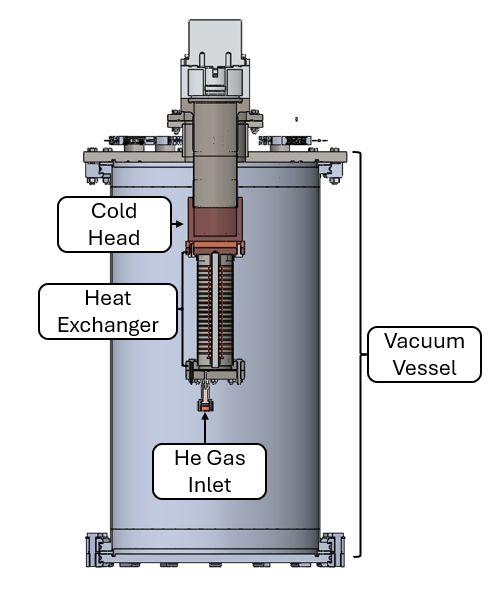}
  \caption{Cool-o-stat}
  \label{fig:sub1}
\end{subfigure}%
\begin{subfigure}{.5\textwidth}
  \centering
  \includegraphics[width=6cm]{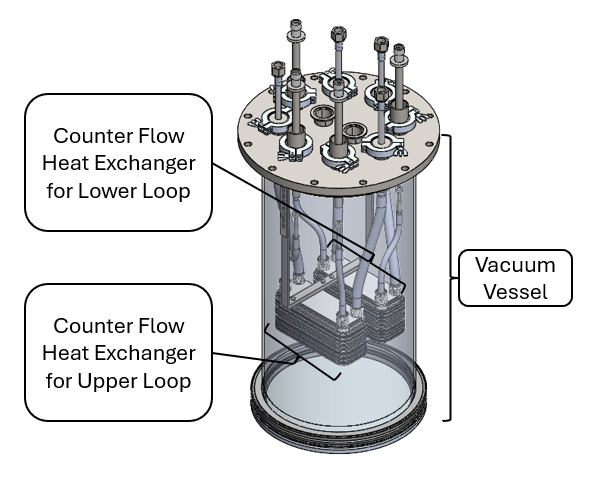}
  \caption{Tardigrade}
  \label{fig:sub2}
\end{subfigure}
\caption{(a) A cut view of the vacuum vessel that houses the cold head and the primary heat exchanger which is mounted to it (``Cool-o-stat"); (b) A view of the vacuum vessel with transparency to show the CFHXs for both of the cooling loops that are housed within (``Tardigrade").}
\label{fig:test}
\end{figure}

Within each loop, a KNF Neuberger N1400.1.2 double diaphragm process pump circulates the helium gas (Fig. \ref{fig:full_system}(a)). With respect to the piping and instrumentation diagram shown in Appendix A, the direction of flow is right to left across the KNF pump (P-001). Just downstream of the outlet of the KNF pump, an MKS~GE250A mass flow controller \footnote{https://www.mks.com/f/ge250a-ge300a-mass-flow-controller} is used to precisely control the flow rate of the helium gas. The flow rate can be changed to make adjustments to the temperature of helium gas. After the flow controller, the gas enters a counter flow heat exchanger that is housed in a vacuum vessel (``Tardigrade,'' Fig. \ref{fig:full_system}(b),Fig. \ref{fig:test}(b)). 

In the counter flow heat exchanger (CFHX) the gas is pre-cooled by thermal contact with cold gas that is returning from the detector system to the KNF pump in the other leg of the CFHX. The gas then exits the vacuum vessel that houses the CFHX via a 6' long vacuum-insulated transfer line and enters a separate vacuum vessel in which a Sumitomo CH-110 cryocooler is mounted \cite{Sumitomo}. The pre-cooled gas flows through the primary heat exchanger which is mounted to the cryocooler (``Cool-o-stat,'' Fig. \ref{fig:full_system}(c), Fig. \ref{fig:test}(a)). See Fig. \ref{fig:test} for a view of the two vacuum vessels and the components they house.

\begin{figure}[h]
  \centering
  \includegraphics[width=0.25\textwidth]{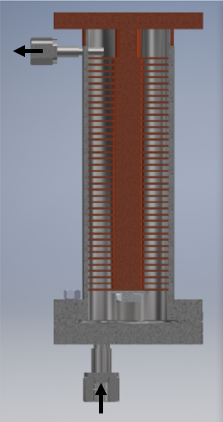}
    \caption{This is the primary heat exchanger. It was custom-built for this detector system. It contains 37 OFHC 101-copper disks separated by OFHC 101-copper spacers. A through hole in the center of the disks and spacers allows them to be stacked over the rod that is concentric with the cylinder that forms the outer shell of the heat exchanger. The disks each contain an additional 1/4" hole that is offset from the center for gas to pass through and neighboring disks are rotated 180$^\circ$ out of phase. Gas enters through the bottom of the heat exchanger.}
  \label{fig:primaryhx}
\end{figure}

As the gas travels through the primary heat exchanger, it passes through a series of oxygen-free high-conductivity (OFHC) 101-copper disks separated by OFHC 101-copper spacers (Fig.~\ref{fig:primaryhx}). The arrangement of plates in the primary heat exchanger maximizes the path length through the heat exchanger and the turbulence of the helium flow, increasing the cooling efficiency of the cold head. Helium exits the heat exchanger at
sub-100 K temperatures. This is the coldest point for the helium gas in the system.

Before it exits the lines within the Cool-o-stat vacuum vessel, the gas passes over an in-line cartridge heater and an in-line platinum resistance temperature detector (RTD) (these can be seen in Fig. \ref{fig:cool_labels}). These components form the essential constituents of a 
proportional–integral–derivative (PID) feedback loop, which can be used to regulate and stabilize the temperature of the helium coolant. The PID loop itself is controlled by a Lakeshore 336 Controller \footnote{https://shop.lakeshore.com/temperature-products/temperature-controllers/336.html} which operates by monitoring the RTD and adjusting the power output of the cartridge heater. The helium is then carried to the detector system by a long vacuum-insulated transfer line (38' for the upper detector and 47' for the lower detector). 

\begin{figure}[htb]
 \centering
  \includegraphics[width=1.0\textwidth]{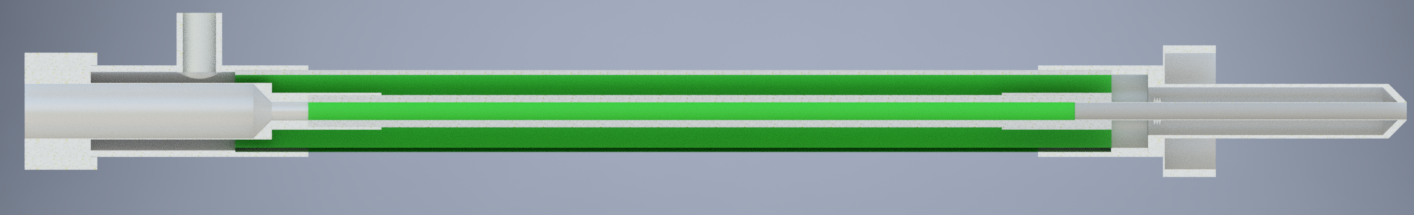}
  \caption{Cross section of a CAD model of a Nab HVFT.} \label{fig:HV_lines}
\end{figure}

\begin{figure}[htb]
  \centering
  \includegraphics[width=0.80\textwidth]{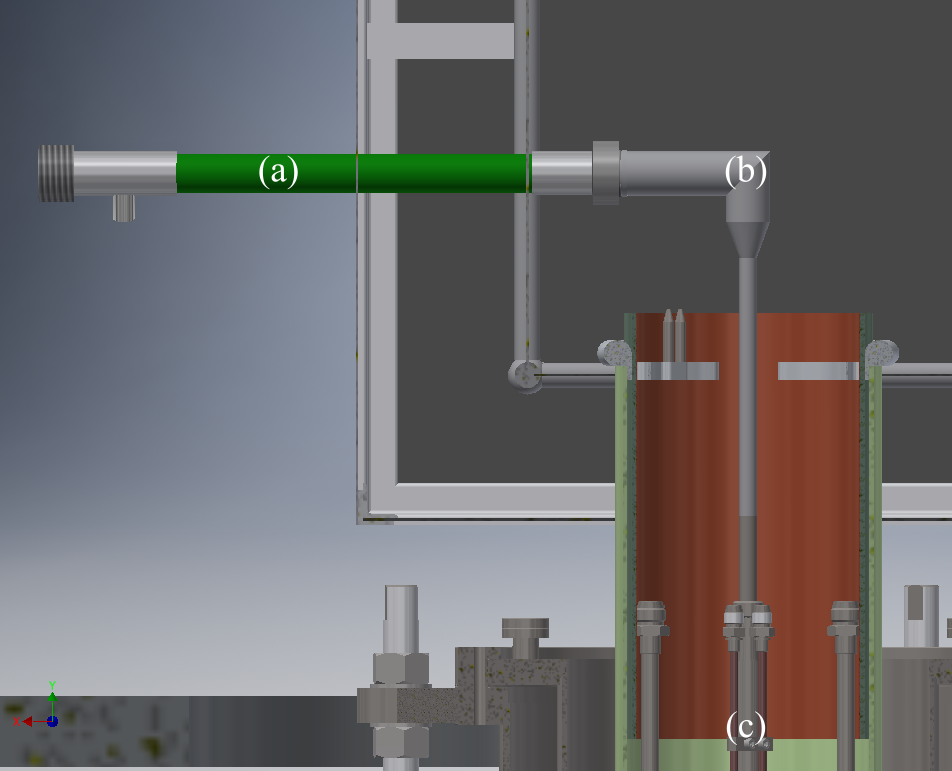}
    \caption{CAD model of the connections between (a) the HVFT, (b) the custom connector, (c) and the detector system cooling lines.}
  \label{fig:HV_FT_Spect}
\end{figure}

The transfer line mates with the specialty component that goes across the high voltage break which we have termed the G-10 High Voltage Feed Through (HVFT) (Fig. \ref{fig:full_system}(d), \ref{fig:HV_lines}, and \ref{fig:HV_FT_Spect}(a)). These custom HVFTs were designed and fabricated to transport the gas into the high voltage cage, across the 30 kV potential drop between the inner and outer cage. To maintain an electrical break across the inner and outer high voltage cages, G-10 was used for both the outer vacuum jacketing and the inner transfer line of the feed through. The G-10 transfer line and vacuum jacketing were epoxied to 3" Linde style Cryofab bayonets. Within the inner high voltage cage, the HVFT mates with the detector mount by means of an additional custom bayonet-style connector.

\begin{figure}[ht]
  \centering
  \includegraphics[width=1.0\textwidth]{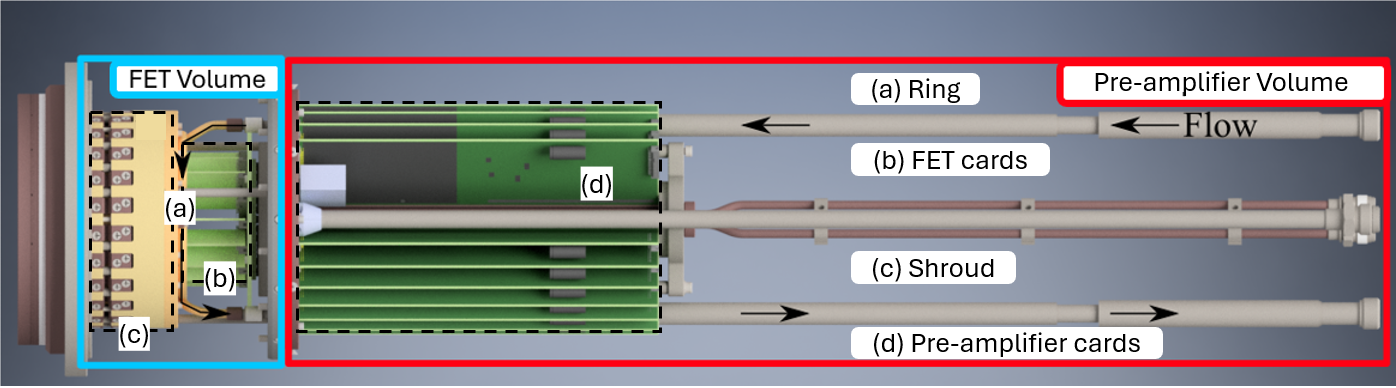}
    \caption{A 3D model of the inner components of a Nab detector system. The pre-amplifier volume is maintained at room temperature and 1 atm. Helium gas enters the mount in the upper line and travels downstream through the mount as indicated by the arrows. The FET volume is evacuated to $\sim 10^{-4}$ Torr. The cooled helium gas circulates around the ring (a).}
  \label{fig:Det_mount}
\end{figure}

This custom bayonet is vacuum-jacketed and it has a right-angle in it which necks down to a lance (Fig. \ref{fig:HV_FT_Spect}(b)) that feeds the gas into the 1/4" inner diameter vacuum-jacketed stainless steel line that is part of the detector system (Fig. \ref{fig:HV_FT_Spect}(c)). The line in the detector system goes through a vacuum break into the FET can. The FET volume is maintained at a vacuum level around a few $10^{-4}$~Torr. Thus, no vacuum jacketing is required for the cooling line in this volume. The cooling line wraps around within the FET volume roughly 1.5 times forming the ``Ring" (Fig.~\ref{fig:Det_mount}(a)). The shroud assembly (Fig.~\ref{fig:Det_mount}(c)) functions as a heat exchanger between the helium gas and the detector and FET cards. The FET volume is directly behind the detector, which is where the cold helium dissipates the heat load on the detector. The gas then exits the FET volume to an identical 1/4"~id stainless line, then cryogenic elbow, G-10 HVFT, and goes to a long vacuum-jacketed return transfer line (39' for the upper detector and 48' for the lower detector). The return line goes to the vacuum vessel that houses the CFHX (Fig.~\ref{fig:full_system}(a) and \ref{fig:test}(b)). Here, the still somewhat cold gas comes into thermal contact with the warm gas leaving the KNF pump in the opposite leg of the CFHX. This warms the helium to above 273~K before it enters the KNF pump where the cycle through the loop repeats. 

The heat load to the FET volume from the room-temperature pre-amplifier volume (Fig.~\ref{fig:Det_mount}) comes through the vacuum feedthrough that separates them. The pre-amplifier volume is maintained at a stable temperature by a water-cooling system \cite{nelsenuky}.

\section{Instrumentation}
\begin{figure}[h]
  \centering
  \includegraphics[width=1.0\textwidth]{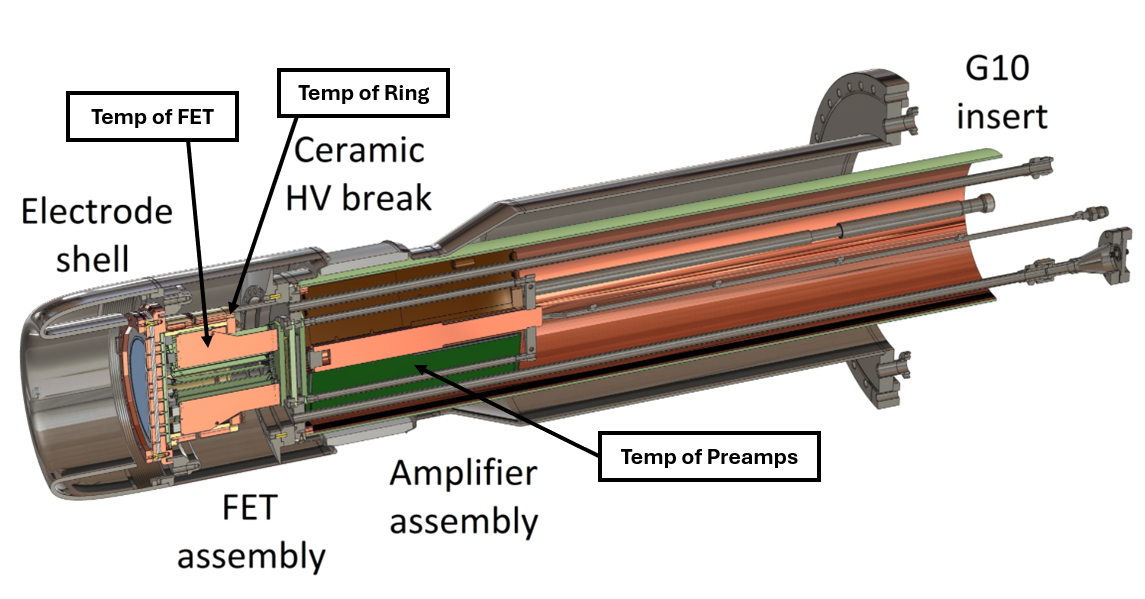}
    \caption{A cut view of the detector system with labels for the approximate locations of temperature sensors.}
  \label{fig:Det_mount_labels}
\end{figure}

The Nab detector cooling system is instrumented with several temperature sensors. There are two in-line platinum RTDs in each loop.
\begin{figure}[!ht]
  \centering
  \includegraphics[width=1.0\textwidth]{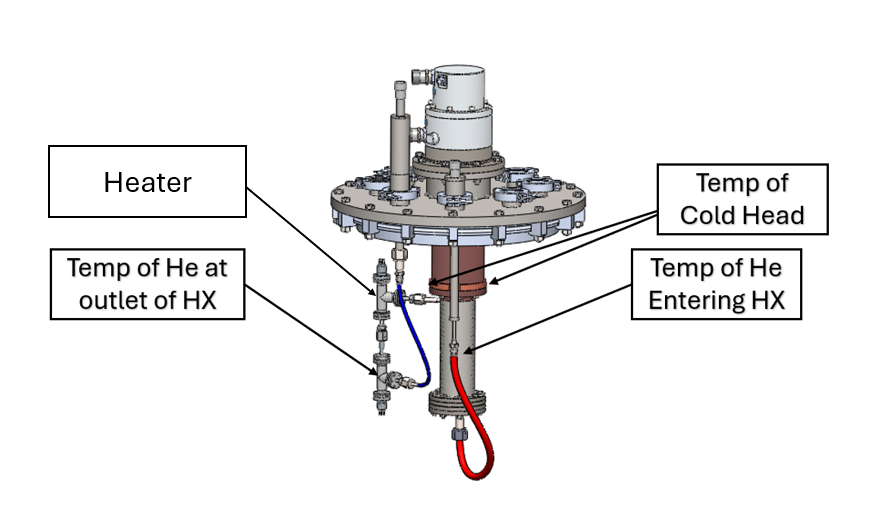}
    \caption{A view of the cold head and custom heat exchanger inside the Cool-o-stat with labels for the approximate locations of temperature sensors and the in-line heater.}
  \label{fig:cool_labels}
\end{figure}
The approximate locations of the sensors within the Cool-o-stat assembly are shown in Fig. \ref{fig:cool_labels}. One is at the inlet to the primary heat exchanger to monitor the temperature of the pre-cooled gas. The second in-line platinum RTD is used to monitor the temperature of the helium leaving the primary heat exchanger that is mounted to the cold head. Each cold head has two temperature sensors mounted to the interface plate that is between it and the primary heat exchanger (Fig.~\ref{fig:cool_labels}). During the commissioning of this system, temperature monitoring for the detector system itself was inconsistent because the detector system was undergoing commissioning in parallel. Thus, different temperature sensor locations are used in the upper and lower mounts to benchmark performance. We cannot put a temperature sensor directly on the surface of the silicon detector, so the temperature in the FET volume for the upper detector and the temperature on the Ring for the lower detector (approximate locations shown in Fig. \ref{fig:Det_mount_labels}) are used to benchmark the performance of the system.

\section{Performance and Discussion}

The figures of merit for the Nab detector cooling system are its ability to operate continuously for a beam cycle, temperature stability, and detector temperatures achieved. The performance of the Nab detector cooling system during the first simultaneous cool down of the upper and lower detector systems which started June of 2023 and ran through November 2023 is presented here in terms of these figures of merit. They are followed with a discussion of heat load calculations for the system. The heat load calculations use data from a more recent cool down where performance of the system was further optimized. 

\textbf{Continuous operation.} The only interruption in operation during the commissioning period described above was due to a KNF pump diaphragm rupture in the lower loop. The resulting downtime was roughly two days. In the future, downtime from similar failures can be reduced by heat curing the thread-lock on the diaphragm retaining plate. Furthermore, we now have a third circulation pump that we can swap into the system to avoid longer outages due to pump failure.

It is important to understand the performance of the KNF Neuberger double diaphragm process pumps to optimize the performance of the cooling system. The rubber diaphragms in the heads of each pump wear out over time. The best solution to this is to replace the diaphragms routinely. Another factor that can increase the likelihood of a diaphragm rupture is operating the system with the inlet pressure of the circulation pump well-above or well-below atmospheric pressure. Either of these operation modes significantly decrease the lifetime of the diaphragms and reed valves within the pump heads. Finally, the circulation pumps will get very hot if they are not actively cooled. We use chilled water to cool them during operation. This is two-fold in its benefits to the system. It lowers the amount of wear on the diaphragms and reed valves, and it also reduces the contribution of the pumps to the heat load on the system. In summary, by periodically replacing the diaphragms, by running the system so that the pump inlet pressure is approximately equal to atmospheric pressure, and by actively cooling the circulation pumps, we can run the system for full beam cycles (approximately 3 to 6 months) uninterrupted.

\begin{figure}[ht]
    \centering
    \includegraphics[width=\textwidth]{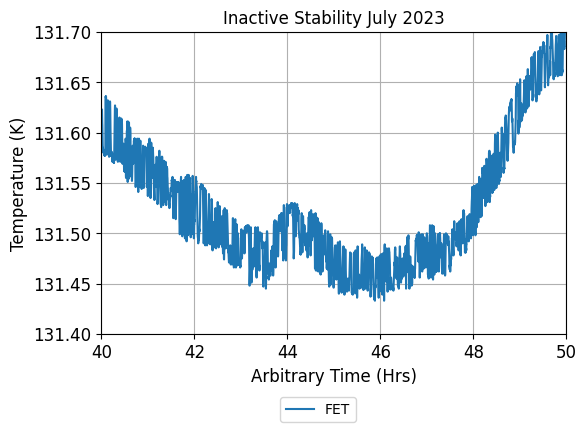}
    \caption{This plot shows the temperature of the upper detector FET volume over a 10 hour period during an inactive period--overnight--while no changes were being made to the detector configuration.}
    \label{fig:stability}
\end{figure}

\textbf{Inactive stability.} The system is capable of achieving $\pm$ 0.2~K inactive stability over a ten hour period (see Fig. \ref{fig:stability}). Inactive stability refers to the temperature stability during a period where the in-line heater is off, the flow rate is constant, and the state of the detector is stable. Under these conditions, the temperature at the FET in the upper detector system varied by less than 0.2~K over a 10 hour long period. The temperature of the detector is sensitive to the amount of power being supplied to the FETs. In addition, active changes in the experiment hall conditions and in the power being supplied to the detector electronics did lead to larger than 0.2~K variations in the temperature. The design goal is to achieve $\pm$ 0.5~K active temperature stabilization, so if we can control external factors we will be able to achieve better than this. Additional work is needed to understand the temperature control repeatability and the stability needs of the system.

\begin{figure}[ht]
    \centering
    \includegraphics[width=\textwidth]{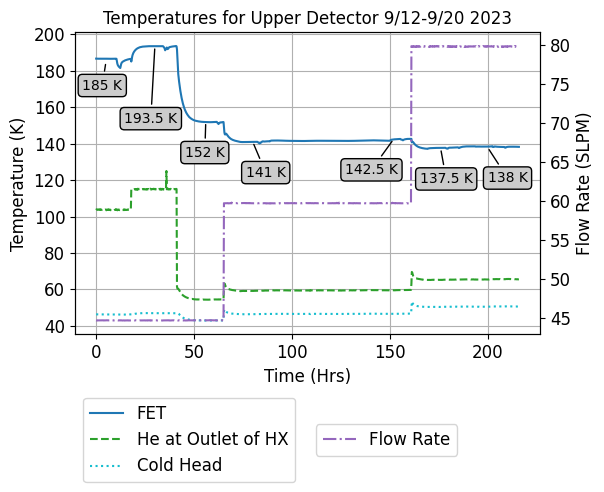}
    \caption{This plot demonstrates the dynamic capabilities of the system. Each stabilization temperature for the upper FET volume is explicitly labeled (grey outlined text). The temperature sensor and flow rate line colors and styles are represented in the key below the plot. The smaller changes in temperature at a constant flow rate were achieved with the in-line heater.}
    \label{fig:dynamics}
\end{figure}

\textbf{Active stability.} With a combination of controlling the flow rate of the helium and the settings on the in-line heater and PID loop, this system is capable of achieving stabilization for the full range of temperatures that it can access. Target temperatures are accessed by dynamically tuning the flow rate and heater settings. It is demonstrated in Fig. \ref{fig:dynamics} that the system is capable of both fine and coarse tuning. This dynamic range enables future studies of temperature effects on the detector performance which are important to understand to reach the precision goals for Nab.
 
\begin{figure}[ht]
    \centering
    \includegraphics[width=\textwidth]{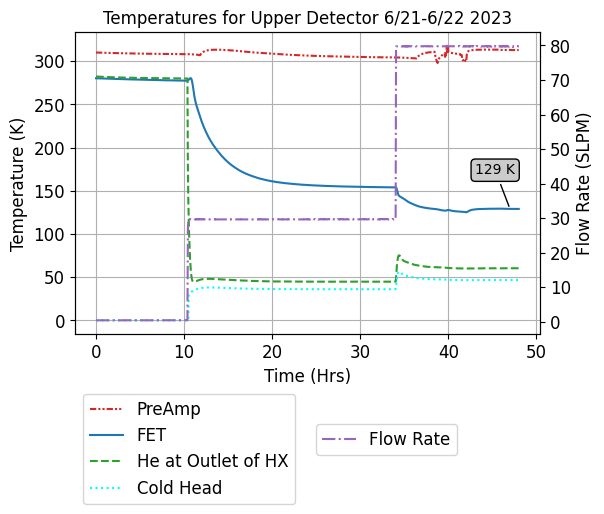}
    \caption{This plot shows the cooling system performance with the upper detector system at the beginning of the cool down in June 2023. The temperature sensor and flow rate line colors and styles are represented in the key below the plot. Near the 10 hour mark, the cool down was started. Around the 34 hour mark, the flow rate was changed from 30 SLPM to 80 SLPM. This pushed the FET volume temperature from just above 150 K to 129 K.}
    \label{fig:tempsup}
\end{figure}

\begin{figure}[h!]
    \centering
    \includegraphics[width=\textwidth]{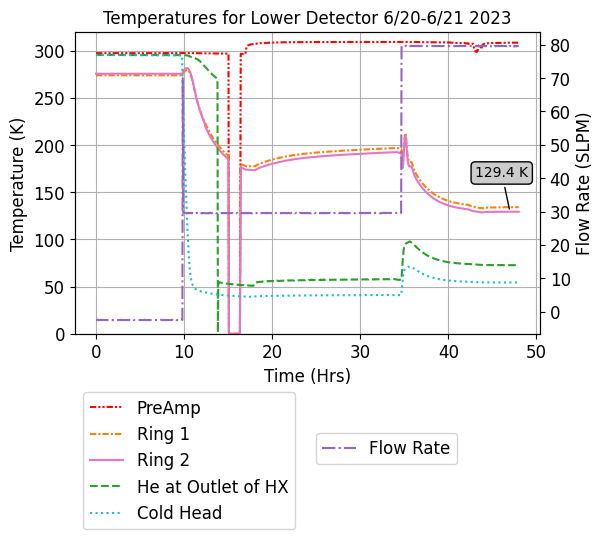}
    \caption{This plot shows the cooling system performance with the lower detector system at the beginning of the cool down in June 2023. The temperature sensor and flow rate line colors and styles are represented in the key below the plot. Near the 10 hour mark, the cool down was started. Around the 15 hour mark there was an interruption in temperature data being collected, but it was restored around the 17 hour mark. Around the 34 hour mark, the flow rate was changed from 30 SLPM to 80 SLPM. This pushed the ring temperature down from close to 200 K to 129.4 K.}
    \label{fig:tempslow}
\end{figure}

\textbf{Temperatures.} As discussed in Section 4, circumstances of commissioning have led to us using two different temperature sensor locations to benchmark performance in the upper and lower cooling loops.  The FET temperature is used for the upper loop and the ``Ring" temperature is used for the lower loop. During the June to November 2023 cool down, the FET temperature in the upper detector system reached 129 K, and the ``Ring 2" temperature for the lower detector system reached 129.4 K (See Fig.\ref{fig:tempsup} and Fig.\ref{fig:tempslow}). This meets the goal for the cooling system of reaching temperatures below 150 K. This cool down was an element of Nab's commissioning data taking. Results from this commissioning effort are shared in Refs. \cite{menu_proceedings,Nabteardrop}.

The table below summarizes the temperatures reached in the June to November 2023 cool down. The values that are used for benchmarking are bolded. (Note: ``heat exchanger" is abbreviated with ``HX" in the table.)

\begin{table}[h!]
\caption[Temperatures at different positions in the system]{Temperatures at different positions in the system during Nab's commissioning.}
\begin{center}
\resizebox{1.0\textwidth}{!}{
\begin{tabular}{ |c|c|c|c|c| } 
 \hline
 Sensor Position & Preamplifier Volume & \textbf{FET Volume} or \textbf{Ring 2} & Outlet of Primary HX & Cold Head\\
 \hline
 Upper & 312.9 K & \textbf{129 K} & 60.3 K & 46.7 K \\ 
 \hline
 Lower & 308.5 K & \textbf{129.4 K} & 73.0 K & 50.4 K \\
 \hline
\end{tabular}}
\end{center}
\label{tab:temps}
\end{table}

It is notable that the performance of this cooling system is improved by pumping on all insulation vacua in situ. During the tests discussed in Section 5 the insulation vacuum for the long return transfer line from the detector system back to the Tardigrade for each loop was not being actively pumped on during the cool down. With all insulation vacua connected to turbo pumps, sharing of turbo pumps for separate volumes was done where possible. The coldest temperature for the lower loop was lowered from 129.4~K to about 115~K, and the upper loop was lowered from 129~K to about 100~K.

\clearpage

\section{Heat Loads}

During the design-phase of the Nab detector cooling system, a series of engineering calculations were done to predict the detector system heat load. To make these estimates, a few assumptions were made: 1) all thermal connections are perfect, 2) radiative cooling from the bore of the magnetic spectrometer is neglected, and 3) heat from the preamplifier assembly is rejected via dry nitrogen or air circulation and the temperature at the cryo feed through is 300 K. With these assumptions and with attributing 3.2 W to the FET electronic heat load, the heat load from the detector system was estimated to be 32.12 W for each loop. 

\begin{figure}[ht]
    \centering
    \includegraphics[width=\textwidth]{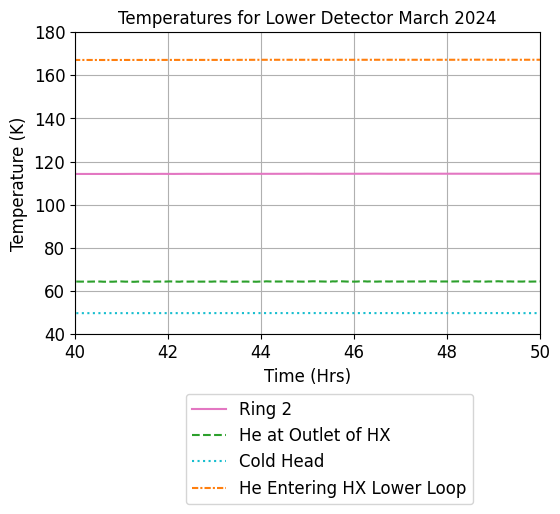}
    \caption{This plot shows the temperatures used for the heat load calculations. The temperature sensor and flow rate line colors and styles are represented in the key below the plot. The temperatures are lower than those in the performance section because post the commissioning run with both detectors we found that by actively pumping on all the insulation vacua the system could achieve significantly improved performance.}
    \label{fig:coldtemps}
\end{figure}

When actively pumping on all of the insulation vacua that are part of the detector cooling system, we are able to achieve colder detector temperatures. Here, we will refer to the values from running the lower cooling loop with the detector system in the room-temperature magnetic spectrometer to make estimates of the heat loads absorbed by the system in practice. To do this, we assume that the heat load from the transfer lines is 0.68~W/m, and that the heat load from each Linde bayonet connection is 2.163~W as specified by Cryofab \cite{cryofab}. In practice, the heat loads may vary depending on the vacuum achieved in the insulation volume of the transfer lines. We used the following expression to balance the heat loads around the loop and come up with the heat dissipated by the cold head and an approximate heat load for the detector system:
\begin{align}
    W = \dot m c \Delta T,
    \label{eq:1}
\end{align}
where $W$ is heat, $\dot m$ is the flow rate, $c$ is specific heat of helium which is assumed to remain relatively constant over the temperature range of interest, and $\Delta T$ is the difference in temperature across the component under consideration. Using Eq.~\ref{eq:1}, the heat removed from the system by the cold head is $\sim 128$~W, and the heat load of the detector system is $\sim 48$~W. The heat removed by the cold head was calculated by using the difference in the temperature of the helium entering the HX and the helium leaving the HX (see Fig.~\ref{fig:coldtemps}). For the detector system 
\clearpage

\noindent heat load, the calculation was more involved. The heat loads from Cryofab for the transfer lines and Linde bayonets were used to extrapolate from the temperature of the helium leaving the HX to the temperature that the gas ``should" be just before it enters the detector mount. The difference in the ``Ring 2" detector system temperature and the estimated temperature of gas entering the detector mount was then used in Eq.~\ref{eq:1} to determine the approximate heat load of the detector system to be 48~W.

\begin{figure}[ht]
    \centering
    \includegraphics[width=\textwidth]{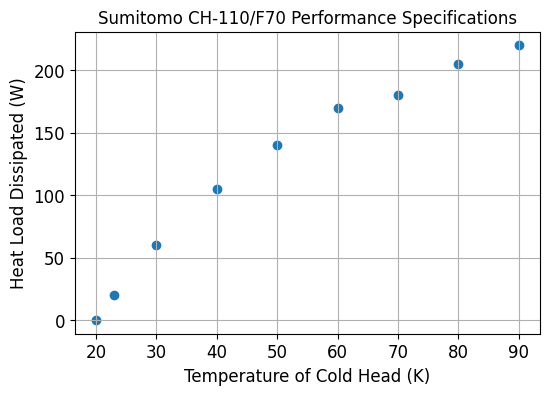}
    \caption{These specifications were provided by Sumitomo \cite{Sumitomo}.}
    \label{fig:coldhead}
\end{figure}

The cooling power curve provided by the Sumitomo CH-110 Cryocooler is shown in Fig. \ref{fig:coldhead}. The cold head was at $\sim$ 50~K for the calculations above (see Fig. \ref{fig:coldtemps}). At this temperature, the cold head should be able to remove 140~W of heat from the system. Comparing this to the 128 W that is removed from the helium as it passes through the heat exchanger indicates that the cool-o-stat assembly has slightly better than 90$\%$ efficiency. At roughly 48 W, the heat load that is calculated for the detector system is $\sim$~50$\%$ higher than the 32~W that was initially estimated. The radiative coupling between the room temperature bore and the detector system will have some contribution to the heat load. Using thermal view factors, and approximating the detector as a disk concentric with a cylinder that represents the bore of the magnet, the heat load that the delta T between the room temperature bore and cold bore will impose on the detector is $\sim$~2 W. Since the Nab experiment runs with the bore cold, the 48~W heat load can be adjusted to 46~W which is $\sim$~44$\%$ higher than the initial estimate. This may be due to thermal connections in the system with efficiencies that are not readily known as well as thermal coupling between the preamplifying electronics and the FETs that is not accounted for.

\section{Conclusion}

A closed-loop, gaseous recirculating helium cooling system for the Nab detectors has been designed, implemented, and commissioned. It has maintained both detectors at well below the 150 K design goal and has achieved inactive temperature stability better than $\pm$ 0.2 K. With proper management of the circulation pumps maintenance, it is capable of running uninterrupted for month-long data-taking periods. The system can stabilize each detector at a wide range of temperatures enabling temperature studies of the detector performance, which are necessary to achieve Nab's precision goals.

\section*{Acknowledgments}
\noindent This research was sponsored by the U.S. Department of Energy (DOE), Office of Science, Office of Nuclear Physics [contracts DE-FG02-03ER41258 and DE-AC05-00OR22725]. This research used resources at the Spallation Neutron Source, a DOE Office of Science User Facility operated by the Oak Ridge National Laboratory. This material is based upon work supported by the U.S. DOE, Office of Science, Office of Workforce Development for Teachers and Scientists, Office of Science Graduate Student Research (SCGSR) program. The SCGSR program is administered by the Oak Ridge Institute for Science and Education for the DOE under contract number DE-SC0014664.

 \includepdf[pagecommand=\section*{Appendix A},scale=0.9,angle=90]{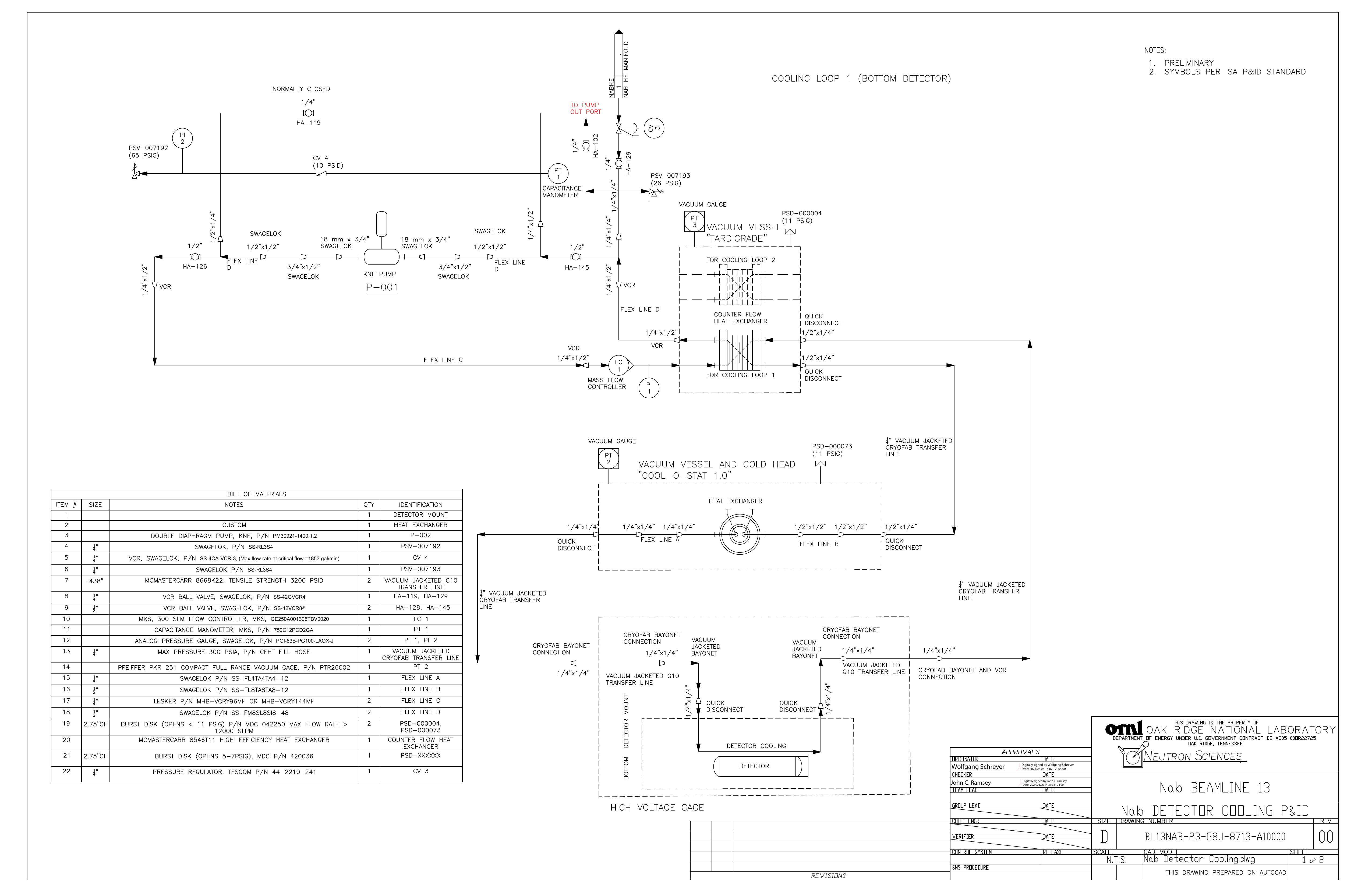}   

\bibliographystyle{plain}
\bibliography{mybibfile}

\end{document}